\documentstyle[11pt,paspconf,epsf]{article}
\def\edcomment#1{\iffalse\marginpar{\raggedright\sl#1\/}\else\relax\fi}

\def\iso#1#2{\hbox{${}^{#2}${\rm #1}}}
\def\he#1{\iso{He}{#1}}
\def\li#1{\iso{Li}{#1}}
\def\be#1{\iso{Be}{#1}}
\def\b1#1{\iso{B}{1#1}}

\def\beq{\begin{equation}}
\def\eeq{\end{equation}}
\def\beqar{\begin{eqnarray}}
\def\eeqar{\end{eqnarray}}

\def\msol{\hbox{$M_{\odot}$}}

\def\lesc{\Lambda_{\rm esc}}
\def\grams{\, {\rm g} \, {\rm cm}^{-2}}

\def\imf{\xi}
\def\dbefe{\hbox{$\omega_{\rm BeFe}$}}

\def\dofe{\hbox{$\omega_{\rm O/Fe}$}}
\def\dij#1#2{\hbox{$\omega_{#1#2}$}}
\def\feh{\hbox{[Fe/H]}}
\def\oh{\hbox{[O/H]}}

\def\ecr{{\cal E}_{\rm CR}}
\def\nsn{{\cal N}_{\rm SN}}
\def\decr{\dot{\cal E}_{\rm CR}}
\def\dnsn{\dot{\cal N}_{\rm SN}}

\def\nupro{$\nu$-process}

\reversemarginpar

\begin{document}

\title{The Galactic Evolution of Beryllium and Boron}

\author{Brian D. Fields}
\affil{Astronomy Department, 1002 W. Green St., University of Illinois,
    Urbana, IL 61801, USA}

\author{Keith A. Olive}
\affil{Theoretical Physics Institute, School of Physics and Astronomy,
University of Minnesota, Minneapolis, MN 55455, USA}

\vskip -3in
\rightline{UMN-TH-1754/99}
\rightline{TPI-MINN-99/16}
\rightline{astro-ph/9903367}
\rightline{March 1999}
\vskip 2.3in

\begin{abstract}
The galactic chemical evolution of Be and B 
provides unique information about 
the origin and history of cosmic rays.  
The available Pop II data demonstrate that Be and B
have a Galactic source,
probably in one or more kinds of spallation processes.
However, the data are not unequivocal about the 
nature of Be and B origin, as encoded in the
primary or secondary (linear or quadratic) scaling with
metallicity.
We summarize a careful analysis of the trends among Be, B, Fe, and O
observations.  We
show that if O/Fe is constant, some other cosmic ray origin or
component is needed.  On
the other hand, if O/Fe is not constant, as recent data suggest, then the
data could indicate a standard cosmic ray origin, wherein the
abundances of cosmic rays scale with those of the ISM.
We suggest future observational tests which will
distinguish several proposed scenarios of LiBeB and cosmic ray origin.
\end{abstract}

\section{Introduction}

Lithium, beryllium, and boron (hereafter, LiBeB) have
a rich nucleosynthetic history.
Indeed, the very low binding energy of the LiBeB isotopes
leaves these nuclides at a thermodynamic disadvantage. 
LiBeB are all burned at fairly low stellar temperatures,
and stars are in fact net destroyers of LiBeB.
Thus these elements have unusual origins.
On the one hand, \li7 is in part primordial,
and both \li7 and \b11 appear to have a component
due neutrino-induced interactions in supernovae.
On the other hand, 
\li6, Be, and \b10 are the ``orphans'' of nucleosynthesis,
made neither in stars nor in the big bang--instead,
their origin lies in the cosmic rays.

Reeves, Fowler, \& Hoyle (1970) noted that
all of LiBeB are produced by cosmic-ray interactions with 
the interstellar medium (ISM).
These authors furthermore showed that the present cosmic-ray flux, traversing
the ISM for the duration of the Galaxy, 
yields LiBeB abundances that are consistent by and large
with observed solar system levels.
The production rate calculations were refined by 
Meneguzzi, Audouze, \& Reeves (1971), who
confirmed the basic success of the mechanism,
but also noted that \li7 seemed to require an extra source
(later, boron isotopic measurements suggested
more \b11 was also required).
Indeed, cosmic ray nucleosynthesis
remained the standard picture for LiBeB production
from its proposal 
until the late 1980's (see also Audouze 1999).

In the past decade, the simple view of LiBeB origin
was enriched (i.e., complicated) by the addition of new data.
The heroic observation of Be (and later, B) in
old, metal poor halo stars
revealed how these elements 
evolve as a function of metallicity.
To appreciate the great impact of these results 
one must first note that 
the {\em expected} LiBeB evolution is readily calculated
within the standard picture of cosmic ray
origin and acceleration.  In this picture,
cosmic rays are accelerated by supernovae,
and have a composition reflecting that of the ISM
(see, e.g., Ellison 1999 and Meyer 1999 and references therein).
This leads to a ``secondary'' dependence of Be and B
on their targets, as follows.
In the early Galaxy, the rate of Be production is dominated
by spallation of cosmic ray protons on interstellar oxygen,
which produces Be atoms at the rate 
$d{\rm Be}/dt \sim {\rm O} \, \sigma \, \Phi_p$.
Each factor in the rate expression has a different dependence on
metallicity.  
Oxygen is produced by supernovae, so ${\rm O} \propto N_{\rm SN}$,
the cumulative number of supernovae.
The spallation cross section $\sigma$ does not depend on the
metallicity, 
while the cosmic ray flux scales as the supernova rate,
$\dot{N}_{\rm SN}$.
Thus we have $d{\rm Be}/dt \propto {\rm O} \, d{\rm O}/dt$
which integrates to ${\rm Be} \propto {\rm O}^2$,
i.e., a logarithmic slope of 2.
This prediction of a quadratic dependence of
spallogenic nuclei on metallicity 
comes about due to the need for target elements
as seeds in the ISM, and thus is also called
a ``secondary'' dependence (vs.\ the  ``primary'' metals).

The BeB observations in Pop II (summarized below 
and in Duncan 1999) emerged
amidst expectations of a quadratic scaling.
Instead, the data indicated a log slope (vs [Fe/H])
close to 1,
for both Be and B--i.e., the data seem to show that Be and B
are primary (again, vs [Fe/H]).  
This result came as a surprise, and has led many
authors to conclude 
that the standard cosmic ray scenario is at best
incomplete or at worst
simply incorrect.  
Proposed explanations include
a new component of accelerated particles, or
a revision of GCR acceleration; these will be discussed below.
On the other hand, it has been emphasized recently that 
the apparently primary nature of Be and B 
implicitly rests on an assumption that
O/Fe is constant in halo stars.
If O/Fe is not constant, as recent observations
suggest (below and in Garcia-Lopez 1999),
then it is possible that standard cosmic ray nucleosynthesis
may yet be revived (Fields \& Olive 1999a,1999b).

Below we will compare the predictions
for different models LiBeB evolution.
Fortunately, current models predict evolutionary
trends with differences which future observations can discriminate.
The data will reveal the nature of LiBeB production
in the early Galaxy.

\section{LiBeB and Metals:  Pop II Trends}

Li elemental and isotopic abundances are
discussed in Olive \& Fields (1999).
Here we concentrate on the Be and B data
in halo stars.

\subsection{Be and B Data}

The past decade has seen 
much progress in BeB abundances in Pop II stars
(Duncan 1999 and reference therein).
{}From heroic first observations, now trends have emerged.
To model LiBeB evolution, one needs accurate abundance data.
In turn, to infer abundances from measured line profiles
requires atmospheric models.
These models can adopt different assumptions,
notably that of LTE vs NLTE, see e.g., Kiselman (1999).
Even within a particular model, the stellar parameter
inputs ($T_{\rm eff}$, gravity, [Fe/H]) can vary
when obtained using different methods.

\begin{table}
\caption{Pop II logarithmic slopes for Be and B versus Fe and O}
\begin{center} \scriptsize
\begin{tabular}{cccccc}
 metal tracer & method & metallicity range & Be slope & B slope & B/Be
slope
\\
\hline
Fe/H & Balmer & $-3 \le \feh \le -1$ & $1.21 \pm 0.12$ & $0.65 \pm 0.11$
&     $-0.18 \pm 0.15$ \\
O/H & Balmer  & $-2.5 \le \oh \le -0.5$ & $1.76 \pm 0.28$ & $1.84 \pm
0.58$ & $-0.81 \pm 0.44$ \\
Fe/H & IRFM & $-3 \le \feh \le -1$ & $1.30 \pm 0.13$ & $0.77 \pm 0.13$
&     $0.01 \pm 0.14$ \\
O/H & IRFM  & $-2.5 \le \oh \le -0.5$ & $1.38 \pm 0.19$ & $1.35 \pm 0.30$
& $0.00 \pm 0.17$ \\
\end{tabular}
\end{center}
\end{table}

To get accurate BeB trends versus metal indicators,
it is essential to use a uniform data set.  
That is, the abundances must be derived using
consistent assumptions about LTE/NLTE, 
and a set of stellar parameters
derived in same way.
In literature, more than one method exists
for determining stellar parameters, giving
qualitatively similar but quantitatively different
results; these differences can obscure BeB trends
if one naively adopts data using more than one method.  
Here, we will present results for data
which uniformly use stellar parameters
derived via (1) the infra-red flux method (IRFM)
and (2) Balmer lines.
For further details, see Fields \& Olive (1999a)
and references therein.

Results appear in Table 1; stellar parameter
techniques as indicated; B is NLTE.
The data are describe by the log slopes,
e.g., \dbefe, defined by
$[{\rm Be}] = \dbefe \, [{\rm Fe/H}] + \hbox{const}$,
where
$[A/B] = \log { (A/B) / (A/B)_\odot }$;
and $[A] = 12 + \log (A/{\rm H})$.
These slopes are fit over the Pop II metallicity
ranges indicated.
In Table 1, we indeed  see that 
the Be-Fe and B-Fe slopes are near 1,
for both sets of stellar parameters
(we will discuss BeB-O
trends below  in \S3).
We also see systematic differences due to 
the choice of stellar parameters:
in Balmer case, the Be and B slopes are not equal, and B slope
less than one, while the   
IRFM gives Be and B slopes consistent with each other,
but different from the Balmer values.  
These difference highlight the importance of 
uniform data sets, and emphasize the need for
a consistent determination of stellar parameters.

\subsection{O and Fe Data}

In the spallation process, the nucleosynthetic
origin of Be and B are directly traced by oxygen rather
than iron.  This distinction is essential to understand
in Pop I, where O/Fe has long been known to decrease with [Fe/H].  
In Pop II, it has commonly been claimed that O/Fe
is constant, in which case the O-Fe distinction is not important
in determining Be and B origin.  
Moreover, different methods (i.e., different lines)
used to determine oxygen abundances in Pop II have 
been reported to give conflicting results.
Thus, iron has been the metallicity indicator of choice,
and iron slopes have been used as indicators
of Be and B origin.

However, recent 
studies of oxygen abundances in Pop II (Garc\'{\i}a-L\'{o}pez 1999; 
Israelian, Garc\'{\i}a-L\'{o}pez, \& Rebolo 1998;
Boesgaard, King, Deliyannis, \& Vogt 1999)
claim agreement among the methods.
Furthermore, these studies 
have shown that O/Fe {\em does} vary significantly.
Namely, O/Fe {\em increases} towards low metallicities.

Following these recent studies,
we allow for changing O/Fe by writing
\beq
[{\rm O}/{\rm Fe}] = \dofe [{\rm Fe}/{\rm H}] + \hbox{const}
\eeq
fit over Pop II metallicities:  $-3 < {\rm [Fe/H]} < -1$.
Israelian et al.\ (1998) find $\dofe = -0.31 \pm 0.11$;
i.e., O/Fe variation seen at the $3\sigma$ level.
Very recently, 
Boesgaard, King, Deliyannis, \& Vogt (1999)
have also reported variations in O/Fe, with
$\dofe = -0.35 \pm 0.03$. 
The two groups' results are completely
consistent with each other, but
quite inconsistent with $\dofe = 0$ (i.e., O $\propto$ Fe).
Variations in O/Fe directly impact the BeB situation,
as we now discuss.

\section{O/Fe and the Phenomenology of BeB Origin}

Motivated by the Israelian et al. (1998) results,
we henceforward will
allow O/Fe to vary in Pop II, and
will explore the consequences of this variation.
Thus, we will put $\dofe \ne 0$, 
which means that
\beq
[{\rm O}/{\rm H}] = [{\rm O}/{\rm Fe}] + [{\rm Fe}/{\rm H}] 
   = (1+\dofe) [{\rm Fe}/{\rm H}] + \hbox{const}
\eeq

Consider the evolution of 
nuclide ${\cal A}\in{\rm LiBeB}$.
Since O/Fe varies, the slopes \dij{\cal A}{\rm O}
and \dij{\cal A}{\rm Fe} will differ.
In particular, up to an additive constant,
$[{\cal A}] = \dij{\cal A}{\rm O} (1+\dofe) [{\rm Fe}/{\rm H}]$
which means that the O and Fe slopes are related by
\beq
\dij{\cal A}{\rm Fe} = \dij{\cal A}{\rm O} (1+\dofe) 
\eeq

Consider the case in which ${\cal A}$ is primary versus O,
so that $\dij{\cal A}{\rm O} \equiv 1$.
Substituting the Israelian et al.\ (1998) O/Fe slope
in eq.\ (3)
gives 
\beq
\dij{\cal A}{\rm Fe} = 1 + \dofe = 0.69 \pm 0.11
\eeq
Note that this is nearly the same as 
B-Fe slope determinations in Table 1.  
Furthermore, we see that a changing O/Fe slope
requires that 
primary elements (vs O) must have slope vs Fe 
{\em less than 1.} 

On the other hand, consider the case 
of ${\cal A}$ secondary versus O,
so that $\dij{\cal A}{\rm O} \equiv 2$.
Now eq.\ (3)
gives 
\beq
\dij{\cal A}{\rm Fe} = 2(1 + \dofe) = 1.38\pm 0.22
\eeq
which is consistent with the Be-Fe slope
determinations in Table 1.
Note also that a secondary slope versus O corresponds to
a slope considerably less than 2 versus Fe.

Finally, if ${\cal A}$ is secondary and another
species ${\cal B}$ primary, then their slopes differ,
and thus their ratio scales with iron according to
a slope 
\beq
\dij{\cal B/A}{\rm ,Fe} = \dij{\cal B}{\rm Fe} - \dij{\cal A}{\rm Fe}
   = -(1+\dofe) = -0.69 \pm 0.11
\eeq
On the other hand, if two elements are both primary (or both
secondary) then their slopes should be the same, and and
their ratio the same.
Thus, ratios of the LiBeB nuclides provide a key test for theories
of nucleosynthesis origin.

We emphasize
that the foregoing analysis is purely phenomenological.
That is, if $\dofe \ne 0$,
this necessarily has an impact on Be and B slopes and inferred evolution,
{\em independent of any model}.
Thus, if variations in halo star O/Fe are confirmed, 
this effect must be taken into account in
{\em any} discussion of LiBeB evolution.
By the same token, if O/Fe were found to be
constant in Pop II (contrary the recent measurements)
then this would establish the need for 
primary Be and B.

\section{Models for Primary LiBeB}

The recent [O/Fe] data
seems to indicate 
that Be has a secondary origin,
as predicted by standard cosmic ray nucleosynthesis
(see \S 5).
On the other hand, the same analysis shows
that B is apparently primary, and thus
requires a production mechanism outside of
standard cosmic ray nucleosynthesis.
Given this, and the present inability of 
the BeBOFe data to definitively discriminate
between primary and secondary scenarios,
it is certainly important to 
study mechanisms for primary LiBeB production.

Several mechanisms for producing primary LiBeB 
use energetic particles and spallation/fusion
of interstellar material.  However, these
processes avoid the standard quadratic scaling with
metallicity by invoking CNO particle compositions
that are constant in time (and thus metallicity),
so that the ratio of energetic CNO/p$\alpha$ remains
fixed, at least roughly (contrary to the standard picture
in which cosmic-ray CNO scales with the ISM metallicity).  
Then at early times,
when the ISM abundances of CNO is down, LiBeB
production is dominated by the ``reverse'' process
of cosmic ray CNO on interstellar H and He.
In this case, the Be production rate is
$d{\rm Be}/dt \sim {\rm H} \, \sigma \, \Phi_{\rm O}$
where the  cross section $\sigma$ and hydrogen abundance H
are time-independent.  The accelerated particle
flux $\Phi_{\rm O}$ 
scales with the supernova rate (or, equivalently, to the star formation
rate), so that $\Phi_{\rm O} \propto \dot{N}_{\rm SN} \propto d{\rm O}/dt$.
Thus we have $d{\rm Be}/dt \propto d{\rm O}/dt$, and 
${\rm Be} \propto {\rm O}$.  Thus LiBeB and
heavy elements are essentially co-produced.

\subsection{Accelerated Particles as Primary Sources}

One proposal for a ``metal-enriched'' accelerated particle
component
invokes a large flux of low-energy ($\la 100$ MeV/nucleon) 
particles dominated by heavy nuclei.  
This suggestion was initially motivated
by reports of $\gamma$-ray line emission from Orion, 
which suggested a low-energy flux dominated by C and O.
Cass\'e, Lehoucq, \& Vangioni-Flam (1995) 
immediately pointed
out that energetic particles of this kind are precisely
what is required to give a primary
LiBeB evolution; as did
Ramaty, Kozlovsky, Lingenfelter (1995).
The Orion $\gamma$-ray detections have since
been retracted (Bloemen 1999),
but even so, the original
Orion data served a useful purpose in that
it stimulated a renewed interest in accelerated particle
interactions outside of standard GCR paradigm.

Indeed, the work on the putative Orion $\gamma$-rays
led to a different but related mechanism:
particle acceleration in superbubbles.
These regions are the seats of intense star formation,
and composed of young stars and rarefied gas which has been enriched 
by massive stellar winds and by supernova explosions.
The stellar winds in these regions produce weak shocks, which
necessarily lead to particle acceleration with
the required enriched composition (Vangioni-Flam et al.\ 1998;
Lemoine, Vangioni-Flam, \& Cass\'e 1998).
As reviewed by Bykov (1999) and Parizot (1999),
the weak shocks lead to steep spectra,
i.e., the fluxes are dominated by low-energy particles.
Thus the particle energies and compositions
are similar to the Orion case.  

While the above scenarios invoke
metal-enriched low-energy particles, a recent proposal
(Ramaty, Kozlovsky, Reeves, Lingenfelter 1996;
Higdon, Lingenfelter, \& Ramaty 1998;
Ramaty \& Lingenfelter 1999)
instead suggests that 
the Galactic cosmic rays themselves
are in fact composed of material 
accelerated from fresh supernova ejecta.
This scenario thus challenges the
standard assumption that cosmic rays
are accelerated out of the ISM.
In this model the particle composition is that
of supernova ejecta, which themselves are
essentially metallicity-independent, and
hence Be and B are primary. 

Finally, we note a suggestion of Tayler (1995)
which has received less attention and has 
not been modeled in any detail.
Tayler notes that a primary Be and B relation 
would arise if star formation and cosmic ray interactions
occur predominantly within objects of globular cluster scales.
The basic idea is that star formation occurs in
a clustered fashion, within giant molecular clouds
of mass $\sim 10^{5-6}\msol$.  The gas in these
clouds will be substantially enriched by supernovae,
with a metal composition that is nearly independent
of the cloud's initial composition (especially in the halo phase).
The fixed target composition thus leads to a primary BeB origin.

\subsection{The Neutrino Process}

While the preceding models all lead to
primary origins for all of the LiBeB nuclides,
one mechanism, the neutrino process ($\nu$-process)
leads is a primary source only
of  \b11 and \li7.
In supernova explosions
thermal neutrinos (of all species) emerge from the hot 
core and traverse the outer layers prior to the propagation
of the shock.  These neutrinos can ``spall''
the nuclei they encounter, 
most likely removing a single nucleon
(Woosley et al.\ 1990; Woosley \& Weaver 1995; 
Hartmann 1999).
Specifically, in the carbon shell, reactions of the form
\iso{C}{12}$(\nu,\nu^{\prime}p)$\b11 
or \iso{C}{12}$(\nu,\nu^{\prime}n)\iso{C}{11}(\beta)$\b11
can produce \b11, but not, as it turns out,
significant amount of \b10 or Be.
In the helium shell,
two-step reactions such as
\he4$(\nu,\nu^{\prime}p)$\iso{H}{3}$(\alpha,\gamma)$\li7
can produce mass-7 (but not significant mass-6).
The yields of these nuclei (Woosley \& Weaver 1995)
are uncertain, but taken at face value,
they 
can be a significant source of \b11 and possibly of \li7 as well.

\subsection{Comparing Primary Models}

By definition, all primary models
predict similar BeB scalings with metallicity.
These models are thus challenging to distinguish observationally. 
However, the models do have real differences in
their predictions over the full
span of metallicities, which enables observational
discrimination with high-quality data.
An instructive case study is provided by 
Vangioni-Flam, Ramaty, Olive, \& Cass\'e (1998),
who compared the predictions of the ``superbubble''
model with the ``direct acceleration'' model.
They found that the two lead to differences in
Be and Be/B evolution which are detectable with 
good Be data at [Fe/H] $\la$ -3.  This result is encouraging
as it shows that observational tests can determine
not only the basic character of LiBeB origins (primary versus
secondary) but can also discriminate among specific
detailed production scenarios.

\section{Models for Standard Cosmic Ray Nucleosynthesis}

Models for standard cosmic ray nucleosynthesis are
important for at least two reasons.
(1) As discussed in \S3, O/Fe observations may suggest
that Be has a purely secondary origin.  If this is so,
then standard GCR nucleosynthesis may be the {\em only}
source of Be, thus making it a crucial part of
the LiBeB discussion.  Furthermore, in Pop II,
the Li isotopes
are primary even in standard GCR, since
$\alpha+\alpha$ fusion dominates and thus the
target He atoms have essentially constant (i.e., mostly primordial)
abundances.  Thus, even if there are other (e.g.,
primary) sources of Li, standard GCR can produce
\li6 and \li7 at comparable or even larger levels.
(2) If indeed GCRs are accelerated out of the ISM,
then this mechanism occurs and is operative throughout
the Galactic history.  Even if other (primary) sources 
contribute to LiBeB, this process must be included
and indeed is significant at late times.
With this in mind, we now review
standard GCR nucleosynthesis and its 
effect on LiBeB chemical evolution.

\subsection{Cosmic Ray Nucleosynthesis}

The details of the GCR model used here are 
presented in Fields \& Olive (1999a);
the formalism and model-dependences
is discussed in detail 
in Fields, Olive, \& Schramm (1994).
To briefly summarize the main points;
LiBeB production rates
are calculated within 
within the 
leaky box model,
following Meneguzzi, Audouze, \& Reeves (1971).
The cosmic ray source spectrum 
is $q(E) \propto (E+m_p)^{-2.7}$ for all particles,
and the source composition at time $t$ are taken to be the
ISM abundances at that epoch.
The propagation model includes losses due to
ionization, nuclear inelastic collisions, and
escape; the latter is parameterized by a constant
escape pathlength $\lesc = 11 \grams$.

\subsection{Galactic Chemical Evolution}

LiBeB production is included in a 
chemical evolution model, described in Fields \& Olive (1998).
Briefly, the model uses the
Woosley \& Weaver (1995) yields for supernovae,
including the \nupro yields.  
For stars in the 1--8 \msol\ range, 
the van den Hoek \& Groenewegen (1997)
yields are adopted.  Stellar lifetimes are accounted for,
i.e., the instantaneous recycling approximation is not made.
Both closed box and galactic wind models were explored,
and each was able to provide a good fit to
the LiBeB results.  Here we focus on the simple
case of the closed box model.  
For the models shown, the IMF is
$\imf \propto m^{-2.65}$, and 
the star formation rate is $\phi \propto M_{\rm gas}$.

The O/Fe ratio as computed in the model
does indeed rise towards low metallicities.  However, while this
is in qualitative agreement with the O/Fe data,
the predicted slope is too shallow, so that
the lowest metallicity points are missed.
Because the model's predicted  O/Fe slopes are too small,
it is unable to test the impact of the observed O/Fe 
slope on LiBeB evolution.  Thus, we have simply adopted 
an Fe evolution such that $[{\rm Fe/H}] = [{\rm O/H}]/(1+\dofe)$,
with $\dofe = -0.31$, the Israelian et al.\ (1998) value.
We still rely on our code to compute the
evolution histories we believe are simple 
i.e., those of \li6BeB and O.
However, we use the observed O/Fe dependence to
get Fe, rather than the {\em ab initio}
Fe yields which are model-dependent (e.g., Type Ia and Type II contributions)
and uncertain (mass cuts).

GCR nucleosynthesis
appears in chemical evolution as a source term for LiBeB.
We take the total cosmic ray flux $\Phi \propto \psi$,
the star formation rate.
The other LiBeB sources included are the
primordial component of \li7, and
the \nupro\ contributions to \b11 and \li7.
We do not include other \li7 sources (e.g., AGB stars),
and thus do not fit the observed Pop I Li abundances.

There are two free parameters for the LiBeB evolution:
(1) an overall normalization to the GCR contributions to
LiBeB, which effectively measures the mean Galactic cosmic ray strength 
today versus that at the formation of the solar system; and
(2) the overall normalization of the \nupro, which
we allow to vary due to uncertainties in the neutrino temperature.
To fix these parameters, we require that 
$\b11/\b10 = (\b11/\b10)_\odot = 4.05$
at $[Fe/H]=0$, which sets the \nupro\ normalization.
The GCR component is scaled using \li6, \be9, and \b10,
which have no other contributions.  Namely, normalization is
to the average of the normalizations of each of these three
to the solar values at [Fe/H] = 0.

\subsection{Results}

\begin{figure}
\plottwo{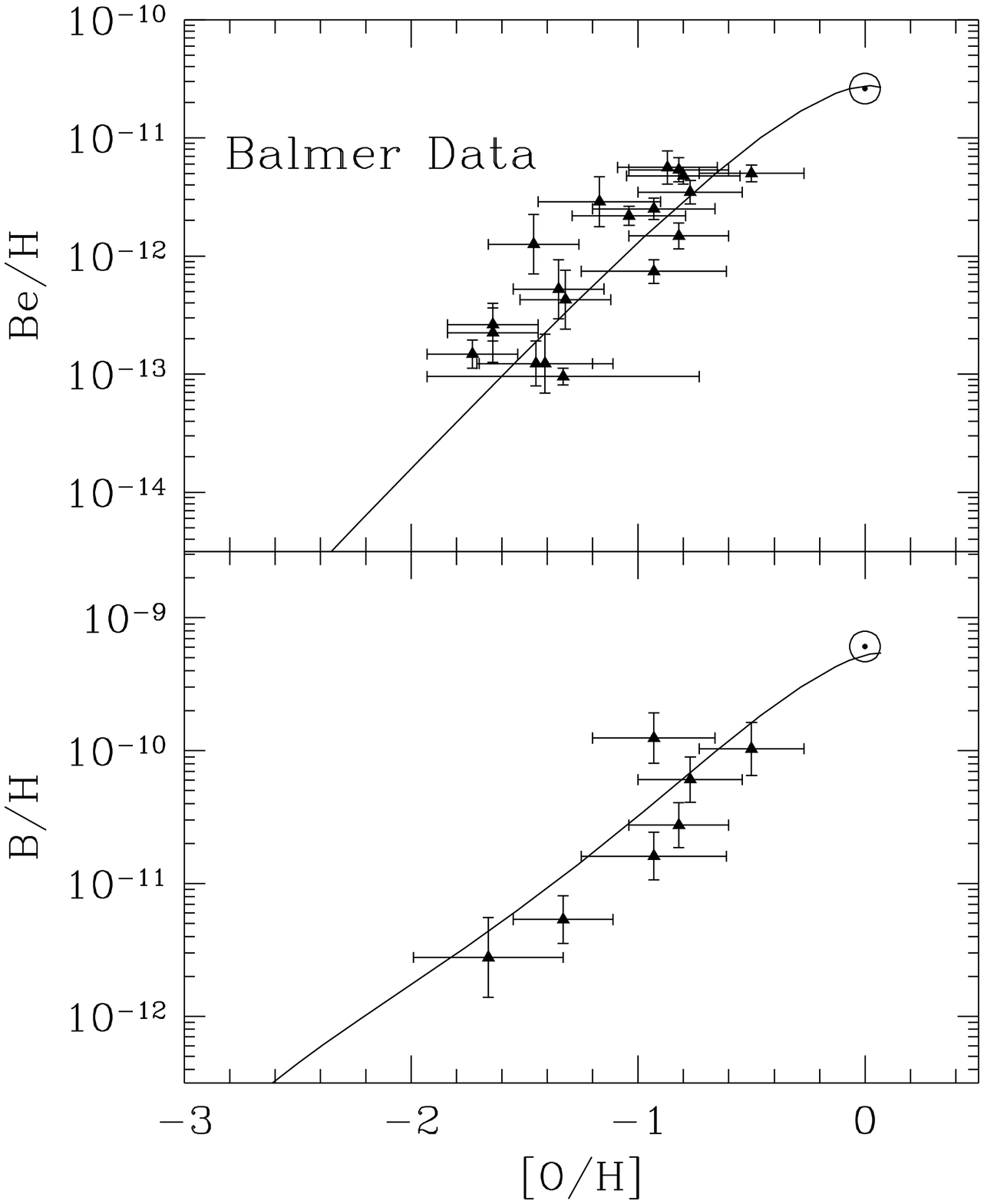}{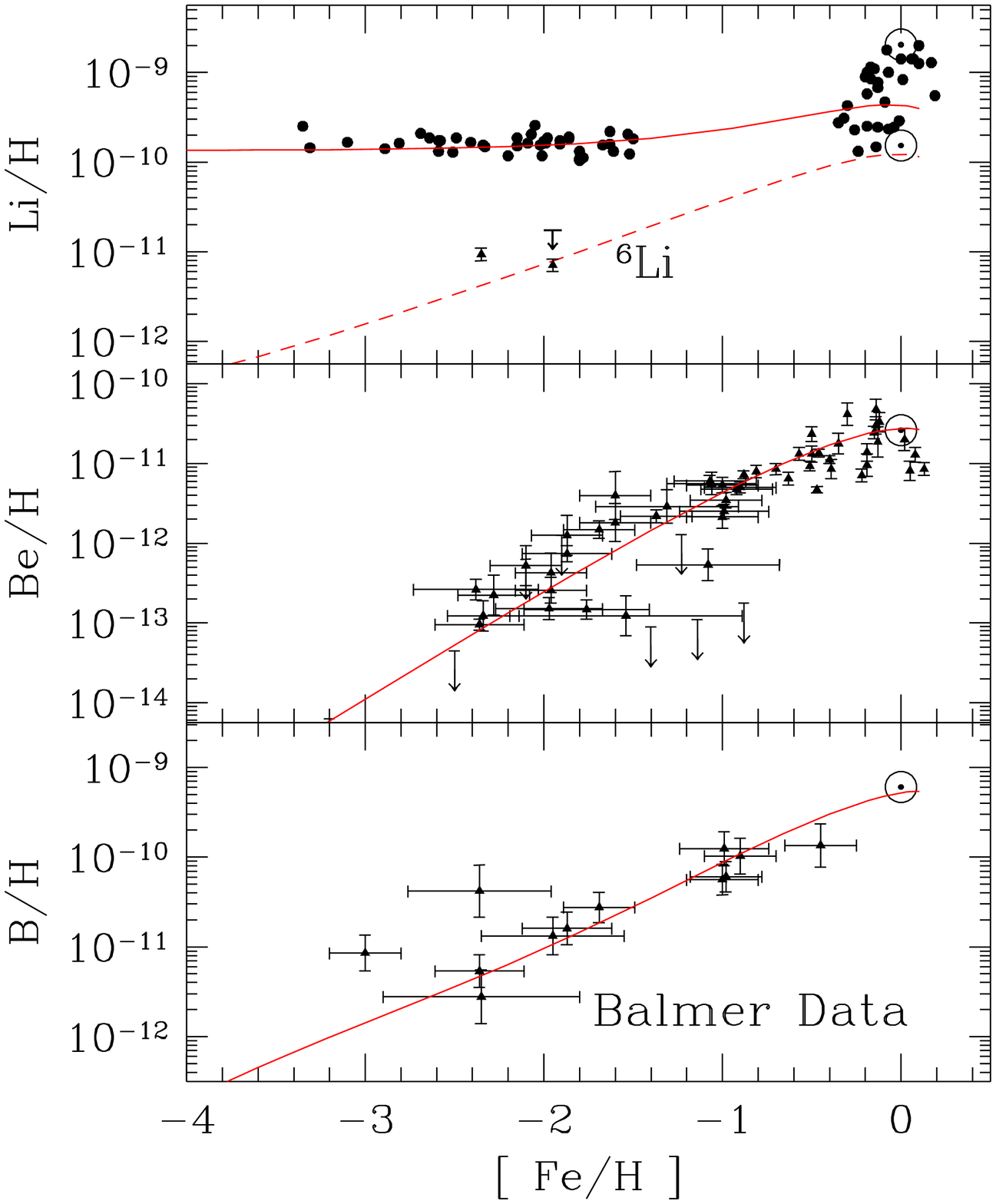} 
\caption{ (a) Results for closed box model with IMF $\imf \propto m^{-2.65}$;
{\em Top:} Be versus O and 
{\em bottom:} B versus O.
Pop II abundance data derived using 
the Balmer set of atmosphere parameters (see text). \\
(b) As in (a),
{\em Top:} Li and \li6, 
{\em middle:} Be, and
{\em bottom:} B versus Fe.
Elemental data are described in the text;
\li6 points described in Olive \& Fields (1999b).}
\end{figure}

Figure 1 shows BeB versus O
and LiBeB versus Fe for the closed box model;
Figure 2 shows B/Be versus Fe.
We see that the models provide a good fit to the
data for the both the abundances and the ratios
This example from a full chemical evolution model
supports the conclusion of our phenomenological analysis 
(\S3):  it is possible that \li6BeB evolution
can be explained solely by a combination of
standard GCR nucleosynthesis and the $\nu$-process.

\begin{figure}
\begin{center}
{\leavevmode
\epsfysize=3.4in 
\epsfbox{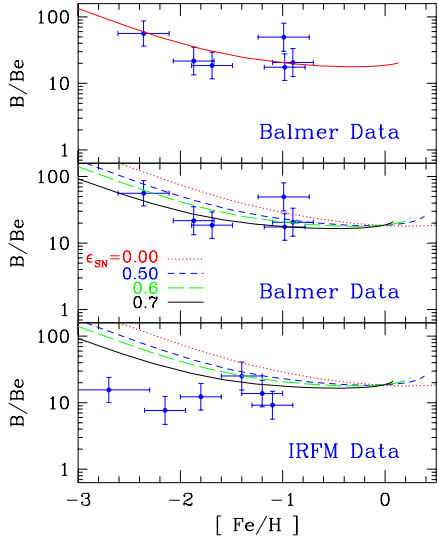}}
\end{center}
\caption{B/Be versus Fe, with 
Pop II abundance data derived using 
the Balmer set of atmosphere parameters (see text).
{\em Top:} closed box model,
{\em middle:} outflow model,
{\em bottom:} outflow model shown with IRFM data.
}
\end{figure}

By calculating the energy production per Be atom
one may link the Be production to the needed energy budget
(see Ramaty \& Lingenfelter 1999 and references therein).
Specifically, one wishes to know the cosmic ray input energy
per supernova,
$\Delta \ecr/\Delta \nsn 
   = \dot{\cal E}_{\rm CR}/\dot{\cal N}_{\rm SN}$
for each epoch $t$, where 
$\ecr$ and $\nsn$ are the aggregate injection
energy and supernova number
over the Galactic history.
The energetics can be related to the observed
Be abundances if one
neglects the astration of Be, and thus 
assumes that the stellar Be/Fe ratio is always
an accurate reflection of the total intergrated
Be and Fe production; this assumption is good in Pop II
but not in Pop I.
Once making this assumption, the energy per supernova is
\beq
\frac{\decr}{\dnsn} = 
   \frac{\decr}{\dot{M}({\rm Be})} \ 
   \frac{\dot{M}({\rm Be})}{\dot{M}({\rm Fe})} \
   \frac{\dot{M}({\rm Fe})}{\dot{M}({\rm O})} \
   \frac{\dot{M}({\rm O})}{\dnsn}
\eeq
One can compute the terms in eq.\ (8)
using (1)
the well-defined ratio of the Be production
rate $\approx \dot{M}(\rm Be)$ to 
$\decr$; (2) the observed Be/Fe data, which
satisfies ${\rm Be/Fe} \propto \dot{M}({\rm Be})/\dot{M}({\rm Fe})$
if the Be-Fe relation is a power law;
(3) the chemical evolution model results for 
$\dot{M}({\rm Fe})$, ${\dot{M}({\rm O})}$, and $\dnsn$.

As noted by Ramaty \& Lingenfelter (1999 and refs.\ therein) 
this analysis points out severe problems
with standard GCR nucleosynthesis
 if one uses the observed Be-Fe trends and takes
O/Fe constant in Pop II.  
In this case, all terms in eq.\ (8) are
constant in metallicity except the first, which
scales as $1/Z$, since this is the abundance of target species.
This implies that the cosmic ray energy input
increases over its present value by a factor of
$10^3$ at $\feh = -3$.  We have implemented
eq.\ (8) in our code for the case of 
constant O/Fe and constant Be/Fe, the
result appears as the short-dashed curve in Figure 3.
While this gives a bad fit to the energetics,
one should recall that one would not consider such
a model anyway, as a constant O/Fe implies
that Be and B are both nearly linear versus O,
in which case a secondary source will of course fail
to fit the data.

The situation changes dramatically, however, if O/Fe is not
constant.  
Both solid curves in Figure 3 use $\dofe=-0.31$.
The upper solid curve has the Balmer value
$\beta=0.2$, leading to an energetic
increase of a factor $\sim 7$; the lower solid curve has
the IRFM $\beta=0.3$, leading to a factor $\sim 3.5$ increase.
These results thus bracket our estimate and give a sense of
the effect of uncertainties in Be-Fe.  To see the effect of the 
whole range of uncertainties, we have plotted in the long-dashed curve
the results for $\beta=0.4$ and $\dofe=-0.4$, both
within $1-\sigma$ of the IRFM slopes. Here we see that
the energetic requirement actually {\em decreases}.  
In any case, we find that within the errors of the
observations, the energetic requirements for our scenario
are not severe, due largely to the effect of changing O/Fe,
and somewhat to the nonzero Be/Fe slope.
We note that different ingredients of this calculation
are subject to uncertainty.
However, for the model we have adopted (the solid curves in Figure 3),
the energetics
are satisfactory, and do not rule out this scenario.

\begin{figure}
\begin{center}
{\leavevmode
\epsfysize=2.5in 
\epsfxsize=2.5in 
\epsfbox{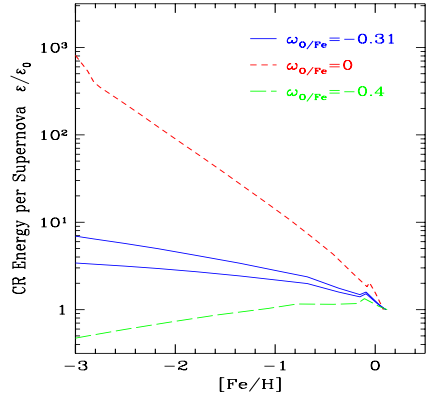}}
\end{center}
\caption{The cosmic ray input energy per supernova. 
The upper solid curve is for a Be-Fe slope
in Pop II of
$\dbefe = 1.3$; the lower solid curve has
$\dbefe = 1.2$;
the short-dashed curve has $\dbefe=1.0$;
and the long dashed curve uses the $1-\sigma$ limit
of $\dbefe = 1.4$.}
\end{figure}

\section{Observational Tests of LiBeB Origin}

It is above all essential to establish the primary versus
secondary character of Be and B.  As noted in \S\S2-3,
the current data are inconclusive on this point,
though the recent O/Fe slopes suggest that Be
is secondary and B primary.  At any rate, when
the basic Be and B origins are clearly established,
further, very accurate data can distinguish
among candidate models,
along the lines suggested by Vangioni-Flam, Ramaty,
Olive, \& Cass\'e (1998).

For the standard GCR model of \S 5,
we predict that all primary to secondary ratios should
vary according to eq.\ (7).  On the other hand,
in primary models all \li6BeB ratios should be roughly constant.
Thus, the most decisive measurements are those that test
whether these key  ratios are seen to vary.
\begin{enumerate}
\item The B/Be ratio.  Current data are sparse,
and also inconclusive due to 
atmosphere uncertainties. 
\item The O/Fe ratio in Pop II.
The O/Fe slope is of course critical
to measure accurately.
Good measurements of oxygen for all stars with Be and B abundances
would also allow a direct determination of the Be-O and B-O slopes
without using Fe as an intermediary.
\item The \li6/Be ratio.  Current data are sparse and uncertain,
but show a rise of \li6/Be 
towards low metallicity, consistent
with standard GCR.  However, more \li6 data is needed,
and it would be particularly useful (however difficult!) to 
have \li6 over a large enough range of [Fe/H] to
see a convincing trend.
\item The \b11/\b10 ratio. Data thus far consists of only
one point, which is uncertain due to possible blending lines.
However, the presence of blending could be tested observationally.
\end{enumerate}
We reiterate that in the analysis of future results,
uniform and consistent stellar atmospheres
are critical for deriving accurate LiBeB-OFe trends.

\section{Conclusions}

Lithium, beryllium, and boron have a unique
nucleosynthetic history which is at the intersection
of cosmology, cosmic rays, and chemical evolution.
Parts of this history are clear:
\li7 is produced cosmologically, while \li6BeB are not,
and the source of \li6BeB is surely spallogenic.
However, the nature of the more detailed history of LiBeB evolution
is uncertain and currently the subject of vigorous debate.
The current Pop II LiBeB data
have enough uncertainty that 
even the basic issue of primary versus secondary origin
has been cast into doubt.  Several well-motivated
models now exist, and can be discriminate by
a larger and better data set.

Thus, the LiBeB field is in the healthy position
of examining fundamental issues LiBeB origin
which bear directly on the question of cosmic ray origin.
What makes the situation exciting is that
these question can be addressed directly by 
observations which are difficult but feasible.
Central issues are the primary versus secondary 
nature of each of the LiBeB elements, and 
the auxiliary but central issue of the O/Fe.
We strongly encourage observations of the kind
described in the previous section, and eagerly
anticipate their results.  
Whatever they may show, the new data will
go far to the establish origin and nature of
cosmic rays in early galaxy.

\acknowledgments
BDF would like to thank the organizers
for a very enjoyable and stimulating meeting.
The work of KAO
was supported in part by DoE grant DE-FG02-94ER-40823 at the University of
Minnesota.

\end{document}